\documentstyle[prl,aps,preprint]{revtex}

\begin{document}

\title{The Limited Scaling Range of Empirical Fractals}

\author{David Avnir,$^{(a)}$ Ofer Biham,$^{(b)}$ Daniel
A. Lidar$^{(b,c)}$ and Ofer Malcai$^{(b)}$}

\address{
$^{(a)}$Institute of Chemistry, the Lise Meitner Minerva Center for
Computational Quantum Chemistry, the Hebrew University of
Jerusalem, Jerusalem 91904, Israel\\
$^{(b)}$Racah Institute of Physics, the Hebrew University of
Jerusalem, Jerusalem 91904, Israel\\
$^{(c)}$Department of Chemistry, University of California, Berkeley, CA
94720, USA}

\maketitle

\section*{}
A notion which has been advanced intensively in the past two decades,
is that fractal geometry describes well the irregular face of
Nature. We were prompted by Marder's recent article in Science \cite{1}, to
comment here on the applicability of this wide-spread notion. Marder
summarizes a simulation study of fractured silicon nitride by Kalia
{\it et. al} \cite{2} which successfully mimics experimental data, and emphasizes the
role of fractal geometry in describing complex-geometry physical
structures, in general. Specifically, the results of Kalia's {\it et. al}
were interpreted as ``showing that this mechanism... leads to fractal
fracture surfaces''. However, upon examining Kalia's results (Fig. 4 in
Ref. \cite{2}) one finds that Marder's statement is based on four exponents,
all of which hold over less than one order of magnitude. We recall
that a fractal object, in the purely mathematical sense, requires
infinitely many orders of magnitude of the power law scaling, and that
a consequent interpretation of experimental results as indicating
fractality requires, ``many'' orders of magnitude. We also recall that,
for instance, in the celebrated fractal Koch flake, one order of
magnitude means about two iterations in the construction and that such
two-iterations Koch curve is not a fractal object. It is our feeling
that Marder, like many others in the scientific community, may have
been swayed by the wide spread image and belief that many-orders
fractality abounds in experimental documentation.

In a recent detailed statistical data analysis we have
shown that this is not the case, at least in the original sense of the
concept \cite{3}: We found that reported experimental fractality in a wide
range of physical systems is typically based on a scaling range which
spans over only 0.5 - 2.0 decades.  The survey was based on all
experimental papers reporting fractal analysis of data which appeared
over a period of seven years in all Physical Review journals
(Phys. Rev. A to E and Phys. Rev. Lett., 1990 - 1996). In these papers
an empirical fractal dimension, $D$, was calculated from various
relations between a property, $P$, and the resolution, $r$, of the general
form

\begin{equation}
P = k \, r^{f(D)}
\end{equation}

\noindent
where $k$ is the prefactor for the power law and the exponent is a
simple function of $D$.  In most cases, fitting the data to Eq.(1) was
done through its linear log-log presentation. Typically, the range of
the linear behavior terminated on both sides either because further
data is not accessible or due to crossover bends. A histogram of the
number of orders of magnitude used to declare fractality, covering all
of the 96 relevant reports, was prepared and is reproduced in
Fig. 1. A clear picture emerges from it: the scaling range of
experimentally declared fractality is extremely limited, centered
around 1.3 orders of magnitude, spanning mainly, as mentioned above,
between 0.5 and 2.0 \cite{4}. This stands in stark contradistinction to the
public image of the status of experimental fractals.

It seems that the most acute questions to be asked in view of
this data are: Is the limited range inherent?; are these limited range
power-law objects, fractal?; and, in fact, is nature describable in
terms of fractality? For reasons detailed next, we are inclined to
propose that the question of fractality is secondary to the benefits
of carrying out a multiple resolution analysis [Eq. (1)]; we believe
these benefits outweigh the perhaps erroneous fractal label. But let
us first begin with the cutoffs of the limited range.

The existence of cutoffs is inherently associated with real
physical objects experimentation. The lower cutoff is dictated
typically either by the basic building block unit (atom, molecule,
microcrystal, small aggregate etc.)  of the system. The upper cutoff
is, at most, of the order of the system size but usually much below
it. It is bounded either by the mechanical strength, or by growth
rates which drop sharply with time, or by the emergence of background
effects such as non-isotropic fields, or by the depletion of
resources. We recall here that many-orders scaling is found in
temporal self-affine trails, but this is a completely different issue:
the time axis can be extended at will.

Do very-limited range power laws represent fractals? Is it
justified to term them as such (5)? It is our view that regardless of
the question of fractality, the more basic question to be asked is: Is
this presentation useful?  The very existence of so many reports by
competent researchers who are well aware of the problematics of
declaring fractality for one order of magnitude experimental results,
suggests that indeed, experimentalists seem to gain from the
resolution analysis and from the fact that the result of such analysis
is often a power law. The usefulness is in the following points:

\begin{itemize}
\item{The power law condenses the description of a complex geometry.}
\item{It allows one to correlate properties and performances of a system to its
structure and to the dynamics of its formation, in a simple way.}
\item{In many instances, the choice is either to use the limited range data, or
to discard it all together and not to have even an approximate picture
of the studied object. Opting for the former can be emphatically
understood.}
\item{Fractal geometry provided a proper language and
symbolism which allowed the front-staging and the legitimization of
studies of ill-defined geometries.}
\end{itemize}

It is important to reiterate however that the ability to
fit data to Eq.(1), does not imply fractality, and that the label
``fractal'' is not needed. So should one refer to such results in terms
of a fractal object? If by ``fractal'' one refers to the original
Mandelbrot teaching of many orders of magnitude, then the data we
collected does not seem to support it in an unequivocal way. If by
``fractal'' one means an object that obeys Eq.(1) over a limited range,
then the use of this label may be acceptable, not only because of its
usefulness, but also because of the following additional reasons:

\begin{itemize}
\item{Interestingly, the sense of self-similarity in irregular objects is
comprehended visually even for a very limited range.}
\item{In some cases,
experimentally derived objects resemble simulative objects obtained
from fractal models.}
\item{The empirical $D$ values for spatial objects
fall in the fractal regime of $0<D<3$.}
\item{And, it may be too late to
make any changes in a terminology which, at this stage, seems to be
deeply rooted in practice. A drift from an original meaning of a
concept is common in science, representing adaptability of the
original ideal definition to realistic restrictions which emerge when
put to practice.}
\end{itemize}

We arrive at our final question: Is then the Geometry of
Nature, Fractal?  Several key processes involving equilibrium critical
phenomena (in magnets, liquids, percolations, phase transitions etc.)
and some non-equilibrium growth models (aggregation), are backed by
intrinsically scale free theories, and lead therefore to power law
scaling behavior on all scales. However, the majority of the data
which was interpreted in terms of fractality in the surveyed
Phys. Rev.  Journals, does not seem to be linked (at least in an
obvious way) to existing models, and in fact, does not have
theoretical backing. Most of the data represents results from
non-equilibrium processes. The common situation is this: An
experimentalist performs a resolution analysis and finds a limited
range power-law, with a $D$ value smaller than the embedding
dimension. Without necessarily resorting to special underlying
mechanistic argumentations, the experimentalist then often selects to
label the object for which she/he finds this power law - ``fractal'' -
this is the Fractal Geometry of Nature.

\section*{FIGURES}
FIG. 1. See page 6 in Ref. \protect\cite{3}:
A histogram of the number of decades of experimentally derived scaling
exponents, which gave rise for declaring the studied system,
fractal \protect\cite{4}.

\end{document}